# At the other end of the sun's path.
# A new interpretation of Machu Picchu.

**Giulio Magli**
**Faculty of Civil Architecture**
**Politecnico di Milano**

*The Inca citadel of Machu Picchu is usually interpreted as a "royal estate" of the Inca ruler Pachacuti. This idea is challenged here by a critical reappraisal of existing sources and a re-analysis of existing evidences. It is shown that such evidences actually point at a quite different interpretation. This interpretation is suggested, on one side, by several clues coming from the urban layout, the interior arrangement of the town, the ancient access ways, the position with respect to the landscape and the cycles of the celestial bodies in Inca times and, on the other side, by a comparison with known information about the Inca pilgrimage center on the Island of the Sun of the Titicaca lake. Altogether, the above mentioned clues lead to propose that Machu Picchu was intentionally planned and built as a pilgrimage center connected with the Inca " cosmovision".*

## 1. Introduction

This paper analyzes one of the most beautiful and enigmatic achievements ever reached by architecture worldwide. It is an ancient Andean town whose original name is unknown; it is anyway famous with the name *Machu Picchu*. Although it may seem strange at a first glance for such a renewed archaeological site, the reason why the town was built, the date at which it was built, the ruler who ordered its construction, the reason why it was abandoned, in a word, the *interpretation* of this place are unknown as well. For reasons we do not know indeed, Machu Picchu was abandoned and forgotten; it was brought again to the attention of the world only with the famous 1911 Hiram Bingham's expedition (see Bingham 1952, or Salazar-Burger 2004 for an up-to-date account). Immediately after its "re-discovery" the site was enveloped by a halo of mystery. Bingham by himself thought it to be the "lost capital" of the last Inca reign, Vilcabamba, an interpretation that we know to be untenable today, and various errors and misunderstandings further contributed to the confusion, such as, for instance, a high over-estimate of the percentage of feminine bones found in the burials, a thing that caused the suspicion that Machu Picchu could have been a sanctuary inhabited by Inca's "Virgins of the Sun". Today, this as well as others, even more unsound theories have been canceled by modern archaeological research (for instance, the true excess of female with respect to male bones is around 1.46 to 1). Modern research also helped, for instance, to clarify the day-by-day life of the inhabitants (see Burger 2004 and references therein). However, contemporary to such important developments, nothing has really come out that may help to explain why and when the Incas built the town (or at least, this is the opinion of the present author, to be substantiated in what follows).
Living without interpretation schemes is extremely difficult in any science (e.g. Physics) and Archaeology is no exception. Therefore, archaeologists have adopted a scheme, a sort of *dogma* on the true meaning of the town: the idea that Machu Picchu has to be identified as one of the "royal estates" of the Inca Pachacuti (Rowe 1990).

What is today customarily called a Inca "royal estate" was a land property, nominally owned by the king and managed by his family clan. A royal estate was typically composed by agricultural lands and "palaces" meant as residences for the ruler and the *elite*. These places were used for amusement (such as hunting) and perhaps also for treating state affairs. A good example of royal estates is Chincero, property of Topa Inca, described in details by the chronicler Betanzos as a property "where to go for recreation" and thoroughly analyzed by Niles (1999). Other important Inca sites have been interpreted as royal estates as well, and in particular Pisac and Ollantaytambo. To the best of the present author's knowledge, however, there is no textual evidence whatsoever showing that Pisac was a royal estate (Pisac is never mentioned in the Spanish chronicles). More convincing is the case of Ollantaytambo (Protzen 1993). This place is in fact associated with the ruler Pachacuti in some Spanish accounts and, in particular, Sarmiento de Gamboa says that the king "took as his own the valley…where he erected some magnificent buildings". As far as Machu Picchu is concerned, it is easy to get the impression from most of the scholarly work that there exist firm textual evidences for the town to be a Pachacuti's royal estate as well. However, as discussed in full details here in Appendix 1, it is *not* so. It is the aim of the present paper to propose a completely different theory regarding the reasons why the town was conceived and built. To "make room" for such an interpretation I am going to use a scientific instrument which is typical of a Physicist's way of thinking: Occam's razor. According to this principle, what is unproved and un-needed not only can, but also should be cut away from any scientific approach to a problem.  I thus attempt to show (Appendix 1) that actually *there is no proof whatsoever*, neither textual nor archaeological, that Machu Picchu was built as a private estate for Pachacuti (or, for what matters, for any other Inca ruler). The interpretation as a royal estate therefore is both unproved and un-needed: Occam's razor allows us to cut it. Further, although I do not exclude that the site may have had *secondary* functions, also the "multi-functional" interpretation adopted by some scholars, who view in Machu Picchu a royal estate but also a sacred and perhaps also administrative center (see e.g. Reinhard 2007, Malville and Ziegler 2007) is refused here. The reason is that this point of view again does not help to explain those characteristics of the site (such the urban project, for instance) which render Machu Picchu *unique* among the Inca towns, and those other characteristics (such as the clear directionality in access and fruition)  which render Machu Picchu *almost* unique, the sole possible comparison being a place whose main function had nothing to do with royal estates or multi-functional centers: the sanctuary on the Island of the Sun (Section 4).

All in all, as a working hypothesis in what follows I will neglect any pre-existing interpretation. This means trying to explore the problem of the meaning of Machu Picchu starting from, and only from, the very beginning: the town itself.

## 2. A  "non-standard" description of Machu Picchu

Machu Picchu  lies at 2400 mts. of altitude, built as a Condor's nest between the two paired peaks (Huayna Picchu to the north and Machu Picchu to the south) which form a sort of peninsula, surrounded on three sides by the gorges of the Urubamba river some 80 Kms north-west of the capital of the Inca empire, Cusco. It would be of course out of the scope of the present paper to give a full description of the site; however, we need a clear idea of the urban layout and of the relationship of the town with the landscape. Here we immediately encounter a curious problem. Indeed, since the city is stretched in an approximate southeast-northwest direction (conform to the general direction of the Machu Picchu-Huayna Picchu ridge), saving of space on paper-sheets usually implies  - starting from the very first Bingham's plan - that north is put lower right. However, this way of mapping the town changes completely the correct perspective in which the place is - and most important *was* - actually *visited* by a newcomer. Further, this unusual orientation makes it difficult to have a glance at the orientation of the buildings with respect to the path of the sun and of the stars, as well as to the relationships of the layout of the town with the cardinal points (and the mountains associated to them). Therefore, although it may seem somewhat a trivial point, I consider *fundamental* for a correct interpretation of the site to refer to a north-on-top map (to this aim I have rotated one of the original maps made by Bingham, see Fig. 1). Further, to understand the layout of Machu Picchu it is important to keep in mind that the complete urban plan was conceived, planned, founded and built from scratch, following the precise will of the planner on the basis of a global unitary project laid out after an accurate and complete survey of the area and, in particular, of the natural rock outcrops which were scattered around. Many such rocks were leveled; others however were "interpreted" in an artistic way, in accordance with that special feeling of the Inca artists for the stone and, in particular, for the shapes of stones that "replicate" natural elements. Huge boulders and existing caves were therefore wonderfully integrated into the project. Each time the planner felt it necessary, the terrain was leveled with the use of a sophisticated technique of superimposed foundation layers. It is thought that as much as 60% of the work involved in the construction of the town is buried in its basements (Wright and Valencia Zagarra 2001). Finally, it is important to notice that the town is deprived of water sources, and therefore a careful project was needed to construct an aqueduct which brings water from a spring located higher on the north flank of Machu Picchu mountain.

The town was abandoned when this huge building program was near to completion, so that some elements - such as for instance the area of the temple of the three windows - were left unfinished, as it was, probably, another magnificent architectural task of the Incas, namely the Sacsahuaman "fortress" in Cusco (Protzen 2004). It is of course difficult to estimate the time employed to bring the site to the present almost-finished state (to the best of my knowledge, nobody has ever tried to figure this time out). In any case, it is hard to believe that from the beginning of the planning to the state visible today - with the inclusion of hundreds of agricultural terraces - less than, say, some tens of years could have lasted (decidedly a long time to wait for a private estate to be ready).

The rigorous, unitary project inspiring the construction of the town remained fully unaltered after the conquest, contrary to what happened to the other Inca settlements. There is, however, a notable exception, the provincial site called Huanuco Pampa (Morris & Thompson 1985). This town was created by the Incas as an administrative center, and as such it has nothing to do with Machu Picchu from a functional point of view. However, it is very useful as a term of comparison in order to gain a better understanding of the inspiring principles of Inca town planning (Hyslop 1990).Huanuco was founded in the central highlands of Chinchaysuyu, at 3700 metres of altitude. The Spaniards gave up rapidly the idea of living at such an unfriendly altitude, and therefore the place was soon abandoned (Fig. 2, right). It exhibits four main "quarters" corresponding roughly to the cardinal directions and surrounding an enormous *plaza* which is empty, except for a squared building which was probably used for ceremonial purposes. The eastern sector of the town – connected to the center by a spectacular east-west alignment of double-jamb doorways - is the unique part built in fine Inca stonework and was certainly devoted to the rulers and to ritual activities.

If we compare the two plans, we immediately see that also Machu Picchu was apparently conceived in a pretty similar, quadripartite way (Fig. 2, left).[1] The division of the town in sectors which will be used here (slightly different to the one usually adopted) is given in Fig. 3; from it we can recognize a *key* difference with Huanuco Pampa: buildings blocks at Machu Picchu are present only in two sectors (I and II). Most of the buildings in these two sectors can be understood trough stylistic analysis of Inca architecture (Gasparini and Margolies 1980, Kendall 1985, Niles 2004). Indeed most of the *Kanchas* (blocks) were clearly conceived as residences of the elite, as shown by the fine stoneworks and the presence of double-jamb doorways. In particular, the block of Sector I which is provided with a private garden and direct access to the first of the gravity fountains (i.e., the purest water) was very probably the private apartment of the ruler when he visited the site. Scattered around in both these residential quarters are, however, many buildings which are clearly conceived for ritual purposes, as it is shown, for instance, by their astronomical alignments (see next Section).

Proceeding anti-hourly we encounter Sector III (the northernmost). This Sector appears to have been intentionally left deprived of buildings. In a sense we can say that it is actually *occupied by the Huayna Picchu mountain*. The central plaza ends at the sides of Huayna Picchu; to the right, near the start of the path ascending to the summit, a visitor encounters the so called *Sacred Rock* complex, a leveled small plaza closed to the west by a huge natural rock which was sculpted in such a way to resemble the profile of Mount Yanatin, visible at the distance. Finally, Sector IV contains, instead of building's blocks, a sequence of very peculiar structures. The sequence of such structures as they are viewed by a newcomer arriving to the town is the following:

1) The gate. Contrary to almost all Inca cities, such as Huanuco Pampa, Machu Picchu was indeed fenced by a wall, with a doorway near the western end. The function of the wall was to create a physical separation between the settlement and the outside without however having a defensive aim; in other words Machu Picchu was not "fortified" (accordingly, the town had no springs or water reservoirs).
2) A "disordered" zone were stone-quarrying activities were carried out. It is perhaps worth noticing that those stones which exhibit regularly spaced drill-bits holes were *not* worked by the Incas but are the results of modern "archaeological" experiments; actually this method – typical of the Romans – was not used by the Incas. They instead used hard hammer-stones and took advantage of natural fissures of the rocks. Inca methods are clearly visible in many other points of the same area.
3) The so called Sacred Plaza. It is a small space open to the western horizon and closed on the other three sides. To the east, in particular, one finds the so-called Temple of the Three Windows, actually a 3-sided building. The windows, a spectacular feat of engineering composed by huge, perfectly dressed polygonal blocks, are located on the east wall facing the central plaza.
4) The so called Intihuatana, a terraced, steep pyramid on the summit of which lies a carved stone of white granite

## 3. Machu Picchu and the landscape

Machu Picchu is wonderfully integrated in the landscape; this holds in a quite "global" sense, including not only the earthly landscape, but also the sky at the time of construction. The word "landscape" is therefore meant here in the broad sense which is of common use today in archaeoastronomy (see e.g. Magli 2009).

To describe the relationship between the town and the landscape, it is worth starting from the analysis given by Reinhard (2007) of the position of the town with respect to the surrounding mountains. Indeed, as is well known, the mountains were a fundamental part of the Inca's *huacas* (shrines).

To visualize the position of Machu Picchu we can draw a cardinally oriented cross with the town at the center, and then follow the four cardinal directions (Fig. 4). In this way we meet four important peaks of the region. First of all, north is identified by the "proprietary" mountain Huayna Picchu, which is, in a sense, a part of the town itself. The summit (2720 m) was - and is - accessed via a spectacular path, essentially a steep flight of steps partly carved in the rock which circles the hill on the west. On the top there are features clearly connected with ritual activities. In particular, a building has its windows looking out toward the Aobamba-Santa Teresa ridge and the recently restudied site of Llactapata, where a corresponding construction focuses the view on Huayna Picchu (Ziegler, Thomson and Malville 2003). The highest point is shaped as a sort of arrow which points due south. In this direction of the cardinal cross, the sight at the horizon is blocked by the Salcantay peak, one of the highest mountain of the region of Cusco (6271 m). Salcantay was certainly revered already in Inca times and is today perceived as the brother of the (slightly higher) Ausangate, the highest peak of the Inca heartland, located east of Cusco.

Important mountains are also located in correspondence to the east and the west of Machu Picchu. Indeed if we take the (slightly elevated) point of view of an observer located on the Intihuatana platform, looking east we can immediately discern one of the two "pyramidal" peaks of Mount Veronica (5750 m). To the west instead, the peaks of the Pumasillo (6075 m) range span the south-western horizon, with the northernmost end of the range located due west.

Interestingly, in accordance with the fact that everything in Inca religion had dual, complimentary aspects, the relationship of Machu Picchu with the mountains is no exception. In fact, the mountains with respect to which the town occupies such a "special place" are "replicated" in many sculpted "image rocks" scattered in the town itself.

Finally, the position of Machu Picchu has certainly to be considered as "special" also with respect to the Urubamba (Vilcanota) river, since - as already mentioned - the river makes a 3-sided turn around the Huayna Picchu-Machu Picchu ridge, which appears therefore as a sort of peninsula marking a neat topographical end to the valley.

To give an overview of the relationships of Machu Picchu with the sky, i.e. of the archaeoastronomy of Machu Picchu, it is necessary to distinguish neatly between the "built", residential part of the town, and therefore the eastern flank (sectors I-II) and the "ceremonial" part located on the western flank (sector IV). Indeed, in sectors I-II a clear interest for accurate celestial observations has been documented in many buildings:[2]

1) The so-called Torreon: a P-shaped building of fine Inca masonry which encircles a shaped stone. The motion of the rising sun near the winter solstice could be followed here trough the shadows cast by a cord affixed at the window (Dearborn and White 1989). Also the Pleiades and Scorpio were probably observed here.
2) The underlying cave (customarily called the Royal Mausoleum) beneath the Torreon, also aligned to the June solstice sunrise (Malville and Ziegler 2007).
3) The cave usually called Intimachay: a natural cavity with a carefully cut tunnel-window which allows a quite precise measurement of the sun rising at the summer solstice (Dearborn, Schreiber and White 1987)
4) A general "solstitial" planning of the whole "royal" sector

People living in the residential sectors had therefore access to devices aimed to follow with good precision the timing of the solstices. Vice-versa, only a generic, rather symbolic interest for the sky is recognizable in the ceremonial sector. Actually, and in spite of several existing claims (regarding e.g. the astronomical function of the Intihuatana stone) to the best of the author's knowledge no clear interest for *precise* astronomical measurements has ever been convincingly documented here. Interestingly, thus, astronomical sightings were aimed at the *rising* of the sun and/or other celestial bodies, while the likely zone where rites were carried out seems to be suited for "popular" observations of phenomena at *setting*. In particular, from the Intihuatana the highest summit of Pumasillo roughly corresponds to the azimuth of the setting sun at the December solstice while the sun at the equinoxes is seen as setting at the northern end of the same range.

Finally, to the possible astronomical observations eventually carried on from Machu Picchu we should add (as noticed by Reinhard 2007) the fact that, since the culmination of the sun between the two zenith passages and that of all the stars of the southern portion of the sky obviously happens due south, the imposing mountain of Salcantay could be used as a useful, distant foresight.

The place were Machu Picchu was built seems therefore to have been chosen because it satisfied an impressive amount of geographical/symbolical requirements. It is, of course, possible - although unlikely - to think that such relationships are due to a chance, or that they were ancillary and not a fundamental key to explain the choice of the site. However, the position of Machu Picchu exhibits also *another* interesting feature. To discuss it, we observe that, together with the cardinal directions, also the directions characterized by a relatively thin "void" of azimuths between 135° and 155° degrees (as measured from Cusco) are of tantamount importance in understanding the complex connections between religion, astronomy, cosmology, and sacred geography among the Incas (Urton 1978, Zuidema 1982a, Magli 2005).

First of all, it has of course to be observed that this "void" characterizes the "preferred" direction of the orography of the geographical region of interest here, which spans some 500 Kms starting from Tiahuanaco, following the Titicaca lake and the course of the rivers up to the Viracocha temple in Ratqui, then Cusco (the Inca heartland) and, finally, bends slightly west to follow the Urubamba valley. Machu Picchu stands, in a sense, at the ideal end of this corridor since the river makes a complete turn around its ridge.

As we shall recall in details later on, this "void" - clearly not by chance - characterized also the "ideal direction" of the Inca cosmological myth. These two aspects merge together with yet another very important one, namely, the fact that the "void" happens to be connected with orientation. Indeed, in the southern hemisphere a "pole star" is *never* available (since precession never brings the south celestial pole sufficiently close to a bright star). As a consequence, the natural way to establish (roughly) the position of the south pole is - and has always been - to follow a bright star or group of stars up to culmination. The most natural choice at the Cusco latitudes is to follow those bright stars of the Milky Way which culminate relatively near the pole, and indeed the principal constellations of the Incas were located along the Milky Way. Among them, particularly relevant for orientation were the stars of the constellations called Crux-Centaurus by today's astronomers, and the dark constellation of the Llama (a dark zone of the Milky Way near Centaurus, perceived in the form of a crouching animal) (Urton 1982). These celestial objects were raising in Inca times (today precession shifted a bit these values) just in correspondence of the "void"; for instance Gamma-crux and Alpha-crux were raising in 1430 AD at azimuths (viewed from Cusco with a flat horizon) around 146° and 152° respectively, while the head of the Llama was raising roughly between 141 and 151 degrees. It has even been suggested that the rising azimuth of Alpha-crux perhaps influenced also the choice of the borderline between the two southern suyus (parts) of the empire (Urton 1978). In *any* case, as we shall see more in deep later on, there is little doubt that *Mayu*, the Milky Way, meant as a huge "double branched" celestial river, played a fundamental role in the Inca "cosmovision".

**4. Machu Picchu as a pilgrimage center**

It has been already proposed by some scholars that Machu Picchu was a *sacred center*, visually and symbolically connected with several other huacas of the region and was therefore a *meta* of pilgrimages. In particular, this idea inspires Reinhard's (2007) book, and similar conceits also appear as part of the "multi-functional" interpretation discussed by Ziegler and Malville (2007). In both these works, however, the "royal estate" theory is anyhow maintained. As mentioned, it is instead the aim of the present work to propose that Machu Picchu not only was a sacred center, but that it was conceived, designed and built *specifically* to be a place of pilgrimage, the last part of it actually *taking place inside the town*. In this respect, at least to the best of the author's knowledge, the interpretation proposed here can be considered as new.[3]

To develop this interpretation it is fundamental to start from a comparison between Machu Picchu and the unique Inca pilgrimage site which is historically well documented and has been the subject of a exhaustive study: the Island of the Sun (Bauer and Stanish 2001).

The Island of the Sun is a rocky islet located near the southern end of Lake Titicaca. For some reasons that nobody has ever dared to investigate, an apparently insignificant, natural rock formation present in the northern part of this island was identified as nothing less than the place of origin of the sun, and therefore of the Incas: this place actually appears, although with different details, in most versions of the Inca cosmological myth. As a consequence, a very important Inca sanctuary was located on the island. The whole site was administrated by the state, and the Incas removed the existing population replacing them with colonists from various parts of the empire; also, a specialized group of women was established with the purpose of serving the sanctuary. The sanctuary area of course included the "sacred rock" from were the sun was born, which was the final objective of the pilgrims. The pilgrimage took place in subsequent stages (Fig. 5):

1) Pilgrims gathered at today's Copacabana, and then sailed to the island from the south.
2) Once landed they followed a path oriented - as the island - in a SE-NW direction. The path ultimately brought them to the most sacred part, which was fenced by a (low) wall.
3) Apparently not all of the pilgrims were allowed to pass the entrance to the northern part of the island; those not admitted could anyway have a look at the rites staying in appositely leveled terraces out of the wall.
4) After the wall, the sacred path passed some other gates and "stations". In particular, near the building today called Mama Ojlia the pilgrims could look at the "footprints of the sun", an area where the exposed bedrock contains natural marks resembling huge footprints.
5) Finally, people gathered in the plaza in front of the Sacred Rock, were they witnessed to rites and, at the time of the winter solstice, at the hierophany of the sun setting between two pillars on a ridge to the northwest.

In Bauer and Stanish words (pag. 247) the layout of the site is described as follows:

"It is not by chance that the final destination of the pilgrims was on the point of land farthest from the mainland, on the northwest side of the island of the Sun. The sanctuary was, like many pilgrimage centers of the world, situated in a remote location that served to emphasize its otherworldliness."

Clearly, the very same words may be applied to Machu Picchu, which is located in a similar way: it suffices to substitute the lake surrounding the Island of the Sun with the Urubamba 3-sides gorge. Actually, as we shall see, the list of similarities is much more longer and impressive than this. Let us, indeed, turn our attention to Machu Picchu again.

There are two main ways to approach Machu Picchu coming from the Inca heartland.

The first one is the world-famous, spectacular route – usually called the *Inca trail* - which detaches from the Urubamba river at Pikillacta (Llactapata), some 30 Kms from the Machu Picchu ridge. It leads to the Inca site of Winay-wayna and then to the so-called Intipunku, a building clearly meant to provide a monumental, controlled access and located south of Machu Picchu, in plain view of the town. The "Inca trail" is indeed certainly Inca, since it is in many points carved in the rocks and passes trough or near several Inca settlements. However, it is a very uneasy route, raising e.g. as much as 4200 meters to cross the "Dead Woman Pass". The travel from Cusco to Machu Picchu following this route certainly took at least four full days to the Inca on his human-transported chair and his retinue. However, especially in the dry season (May to September, which is also the season during which Machu Picchu is supposed to have been visited by the Inca) there is another, much easier and natural way from Cusco to Machu Picchu: just follow the Urubamba valley up to the Machu Picchu ridge and then ascend up to Winay-wayna and the town. This path may become dangerous or even impassable in case of heavy rains, but in the dry season it allowed a 3-days long, quiet travel from Cusco. Also this one obviously was an "Inca trail", and indeed several Inca ruins are scattered near the river (e.g. Torontoy). Interestingly, a further Inca path ascending directly from the river to the town has been recently found in the deep forest on the east flank of the Machu Picchu ridge (Valencia Zagarra 2004). Also this path is endowed with artistic fountains and resting spaces, to confirm its "cerimonial", besides practical, function.

All in all, it appears that the "classical" Inca trail was conceived mainly as a *ceremonial* route, not as a functional one, at least in the dry season. Its very existence contributes to cast serious doubts on the royal estate theory: indeed the ruler traveling to Machu Picchu to spend the winter (dry) season there did not need such a long, uneasy way to arrive to his (anyway very far) estate.

Let us now re-join the path of the pilgrims approaching Machu Picchu from Intipunku. All the roads (including that coming from north-west) meet at the so-called upper Agricultural Area outside the main gate. Here a huge *kallanca*, Machu Picchu's largest building, is located. Clear traces of ritual activities are present here, in particular a "replica stone" which, as does the sacred rock of the town, replicates Cerro Yanatin. In the area there is also a number of piles of small stones of different nature, and at least some were evidently carried to the site from distant places; for instance, a number are rounded river rocks. Probably, thus, at least part of such stones are offerings left upon arrival. The whole area presents clear similarities with the area located outside of the innermost sanctuary on the Island of the Sun as it offers an obstructed view on the western sector of the town and the plaza. We can thus speculate that, exactly as it occurred on the Island on the Sun, people who were not admitted further could anyway look at the events taking place in the town by looking from the platform and the terraces. Curiously, the sector of the Sacred Rock is not visible from here, and perhaps for this reason a "replica" of the same monument was created in this place.

Continuing their way down, the pilgrims eventually reached the gate of the town. The town itself was thus characterized as a *closed space*, and this is unusual for the Incas (actually at Machu Picchu we encounter *most of the few* examples of Inca stone doorways which have devices - rings and recesses - meant for closing them with wooden lintels).

It is anyway clear that the fencing wall was not meant for defense purposes, since it would have been exceedingly easy for any enemy to block the water channel (which comes in the town from outside) with just a few stones and in Machu Picchu there are no water reservoirs. The aim of the wall is therefore to furnish a controlled access and, consequently, to stress the separation of what is inside from what remains outside.

The admitted visitors perhaps left their ritual offerings just near the entrance wall, since many peculiar stone pebbles (mainly of obsidian) have been recovered there. Then, people were fronted by a corridor (service buildings such as stables and magazines are located on the left of the entrance but are separated by a wall) with the imposing view of the Huayna Picchu mountain, the likely final meta of the pilgrimage, just in front of them.

Today, at this point most tourists turn right, visiting the "residential sectors" first. However, in ancient times these sectors - which begin with the Royal residence - were very likely closed to public; therefore, a person entering would have had by necessity to proceed straight in the western sector encountering the sequence of structures we became familiar with in the preceding section (or looking at them from the central plaza): first, the so-called quarry; second, the temple of the three windows and, finally, the Intihuatana platform. But why?

It is my aim here to maintain that sector IV in Machu Picchu, oriented and "scheduled" towards north and Huayna Picchu, was conceived as an image, a *replica* of the path followed by the first Incas in the cosmological myth. This myth has been recounted, although sometimes with different details, in many chronicles. Is starts at the Island of the Sun, from were the first Incas travel underneath the earth following the "void" (SE-NW) direction. They emerge in a place called Tampu-tocco. According to Sarmiento de Gamboa and others, this name means "the house of windows" because "it is certain that in this hill there are three windows" and the first Incas came out from one of these windows. Tampu-tocco is located south of Cusco: the Incas now travel therefore to the north, up to the summit of the Huanacauri hill, where one of them is turned into stone becoming a fundamental huaca of the future empire before arriving in the Cusco valley.

There is, at least in my view, little doubt that the key elements of the myth find a close correspondence in the structures of Sector IV. In fact, a newcomer encounters in succession the three main elements of the myth, symbolized respectively by:

1) The quarry. Although it is a true quarrying area, it is also a zone intentionally left in "disorder" located very near the "royal" and the "ceremonial" sectors. It shows signs of ritual activities as well, such as carvings of serpents on the rocks. It is thus tempting to associate it with Pachamama and to think that it was somehow connected with an image of the underground travel of the first Incas.[4]
2) The sacred plaza. Here the temple of the three windows is to be associated with Tampu-Tocco.
3) The Intihuatana pyramid. It might have been conceived as a replica of the Huanacauri hill, the Intihuatana itself resembling in shape the sacred mountain Huayna Picchu located at the end of the path (Reinhard 2007) as well as the sacred stone-huaca which was located on Huanacauri.

It should be noted that, inspired precisely by the Temple of the Three Windows, Hiram Bingham proposed that the town had to be *identified* with Tampu-tocco. It is *not* my intention here to revive his theory. In fact, there is no doubt on the identification of the Huanacauri hill as one of the most important huacas of the Cusco ceque system (Bauer 1998) and actually most chronicles associate Tampu-tocco with the Pacariqtambo hills south of Cusco (probably the imposing Inca site of Maukallakta was built there to recall the mythical events, see Bauer and Stanish 2001). The idea here is rather that Machu Picchu was conceived, as many other sanctuaries around the world, as a powerful and tangible replica of the holy places of the myth. It is in fact well known that sanctuaries are often built to offer "replicas" related to the same sacred event and/or place in different sites and at different stages of monumentality (think for instance to the various "copies of the Holy Land" constructed in Europe during the Middle Age and the Renaissance). It is an easy process to render such places sacred as well, for instance with the presence of holy images, relics, or oracles.[5] In this connection, it may be noticed the following passage appearing in the Bernabè' Cobo ([1653], 1983) chronicle:

"The Incas had founded a town on the site of Pacariqtambo, and they built on it, in order to make it famous, a magnificent royal palace with a splendid temple. The ruins of this palace and temple remain even today, and in them some stone statues and idols are seen. At the entrance of that famous cave at Pacariqtambo there is a carefully cut stone window in memory of the time when Manco Capac left trough it."

This passage is somewhat strange. First of all, it hardly fits with what is visible - at least today - in Pacariqtambo, while it could be easily referred to Machu Picchu. Further, it gives the impression that the author is speaking about a place where he has never been. It is actually difficult to believe that "statues and idols" could have been left on a site so near to Cusco and known to the Spaniards at the time Cobo was writing.

The connection between "power and replica" is especially true for those sanctuaries which were aimed at the foundation of the temporal power in Durkheim's (1912) sense, such as, without doubts, were those administered by the Inca state. Again, the best known example is that on the Island of the Sun, where cyclical ritual activities took place at scheduled times during the year, with the culmination at the winter solstice hierophany. It is thus tempting to attribute a similar role and function also at Machu Picchu, and to conclude that the place was administered by a dedicated group of "priest-astronomers" (*amautas*) in charge of "controlling" the whole course of the *pacha* - the sacred "space and time" of the Incas - as shown by the astronomical alignments of the buildings of the residential sectors (which include *both* solstices).

Finally, although only at the level of what may be simply an interesting coincidence, it can be noted that there are impressive similarities between the NW sector of Machu Picchu and some 17 century drawings appearing in the so-called Miccinelli Documents (Laurencich-Minelli 2006, 2007) and in the Poma de Ayala Chronicle. Actually, as already noticed several times in the literature (see e.g. again Reinhard 2007) the representation of the Huanacauri idol in the rites of the Inca month of March depicted in this document closely resembles the Intihuatana (Fig. 6, right). Of course, many huacas of this type existed before the conquest; however, the resemblance becomes more impressive if we consider also the Poma representations of Tampu-tocco (Fig. 6, left). This place is indeed shown with its three windows at the base of the Huanacauri hill. Perhaps the choice of the month is not casual as well: March is the unique case in which the Inca name of the month mentioned by Poma exhibits the suffix -*pacha* (Laurencich-Minelli and Magli 2010). The equinox was a moment at which the sun (hanan) and the moon (hurin - thought of as representing the dark hours) were equilibrated having the same "strength"; in this month sacrifices to Viracocha were made and, according to ethnographical data, the full moon before the equinox was the time of a feast devoted to the hanan deads, which in turn were associated with the summit of the mountains.

## 5. Discussion.

As explained in details in Appendix 1, any interpretation of Machu Picchu is doomed to remain speculative unless we eventually find out for sure how the town was called by the Incas and discover written texts mentioning both the town and its function in an explicit way. Clearly, however, the interpretation proposed in the present paper would be strengthen if it may help to explain why the town was built in such an accurately chosen, special position. We are thus lead to investigate if it is possible to individuate a link between topography on the earth and celestial cycles in the sky at the times of the Incas.

As a general observation, it *must* be noticed that this kind of issues are extremely delicate in Archaeoastronomy in general. Actually, very few of the proposed examples worldwide of monuments or human-built landscapes associated with "terrestrial images" of the sky are securely proved (see Magli 2009 for a complete discussion). However, in the case of the Incas we have a quite solid starting point which relies on the already mentioned Urton's (1982) fieldwork with contemporary Misminay people. This research has, in particular, put in evidence the fundamental role played by the Milky Way. In the centuries before the Incas, precession brought the solstitial points near the intersections between our galaxy and the ecliptic (Fig. 7). As a consequence, the position of the sun with respect to the Milky Way could be used to estimate the times of the solstices, which were, in turn, fundamental dates among the yearly Inca rituals.[6] The bright arc of our galaxy in the sky was perceived as a "celestial river" having a "terrestrial counterpart" in the Vilcanota which, in Urton's words, is "equated with the Milky Way". This is, therefore, a hint to the possibility that Machu Picchu, located in such a special position with respect to the Vilcanota river, may have had a special significance connected with the sky as well. Actually, written documentations of two Inca pilgrimages along the course of the Vilcanota exists in the work of the chronicler Cristobal De Molina. The first (studied in details by Zuidema 1982b) moved south of east from Cusco and was connected with the cosmological myth (this pilgrimage was with all probabilities a Capac-Cocha one, namely, ending with a human sacrifice). It occurred at the *winter* solstice and was directed "toward the place where the sun was born" (Fig. 8). The final station was a huaca located on the summit of the Vilcanota hill, some 150 Kms aerial from the capital at the border between the Vilcanota and the Titicaca watershed. Coming back, the priests followed the course of the river bringing offerings to other huacas. The second pilgrimage was instead directed north of west, i.e. in the opposite direction, and occurred a few weeks after the *summer* solstice. During the night, illuminated by torches, runners started from Cusco and reached a bridge in Ollantaytambo, where Coca leaves were offered to the river.

In both cases, a fundamental role was played by three key elements: the opposite solstice timing, the opposite "void" direction, and the Vilcanota river alongside it. The two pilgrimages took thus place on the earth, along the terrestrial image of the Milky Way, towards two "ends" which - due to the timing - were clearly connected with the two "ends" of the sun path at the horizon, the journey of our star throughout the fixed stars during the year. Of course, the path of the sun is the Ecliptic, not the Milky Way, but the start and the end points (the winter and the summer solstice) were both located at the crossroads between the two. Interestingly, the southern pilgrimage ended at the sources of the river, but still far from the "true" place of birth of the sun, the Island of the Sun. Perhaps a similar role was played by Machu Picchu with respect to Ollantaytambo in the case of the northern pilgrimage; actually, also the "inter-cardinal" position of the town with respect to the landscape recalls the presence in the sky of cruciform star-to-star Inca constellations, which were concentrated near the two solstitial points (see again Urton 1982, and Fig. 7).

All in all, the hints of different nature and origin presented in this paper seem to point *all* in the same direction. They suggest that Machu Picchu could have been conceived and built as the ideal - *hanan* - counterpart of the Island of the Sun, in accordance with the duality of the sacred which is typical of the Andean world.[7]

Machu Picchu was indeed located at the ideal, opposite crossroad between the terrestrial and the celestial rivers: the other end of the sun's path.

**Acknowledgments**

**Appendix 1. Review of the evidences for the "royal estate" theory.**

As mentioned in the introduction, a Inca "royal estate" was a land property, nominally owned by the king and managed by his family clan, endowed with agricultural lands and residences for the ruler and the *elite*. The very same interpretation has been repeatedatly proposed for Machu Picchu, and it is the aim of this appendix to give a hopefully complete overview of the evidences - or better, of the almost complete lack of evidences - for this idea.

First of all, to gain any textual information on an ancient place we should know how it was called. As a matter of fact however, we cannot be sure about how the town was called by the Incas. Indeed, we only know that it lays between two paired peaks which were called by inhabitants of the area - at the moment of Bingham "discovery" - with the names *Machu Picchu*, the old peak, and *Huayna Picchu,* the young one. These names were reported to Bingham and he decided to call the town as one of the peaks. However, a town with this name (or, for what matters, with the similar name Picho, see below) is *never* mentioned in the Spanish historical chronicles (at least to the best of the author's knowledge). Of course, although being often confused and/or captious, the chronicles are by far the fundamental written sources of information about Inca life and history. Therefore, *any* attempt to identify the role and the meaning of the town cannot be based on historical sources, and is, as a consequence and by necessity, admittedly speculative (this holds of course for the proposals of the present paper as well). Having said that, we can proceed to analyze the "royal estate" hypothesis. This hypothesis was formulated by Rowe (1990) starting from the fact that a place called Picho (Picchu) *is* mentioned in some 16[th] century documents. In particular, it appears in the account of a travel to Vilcabamba. In this manuscript the author, Diego Rodriguez de Figueroa, cites Picho - apparently *without* having visited it - as a place located on a certain path between Condormarca and Tambo. The same place is also mentioned in a legal document of 1562, first discovered by Glave and Remy (1983) where the existence of a Picho *cacique* (local chief) farming Coca is also cited. Basing on these evidences, Rowe proposed that Bingham or his informants made confusion between the true name of the town - which he thinks was originally called Picho - and the nearby mountain.

This interpretation looks reasonable, although it is, of course, difficult to be *certain* of its validity for a series of reasons. First of all, the name Picchu (peak) is *exceedingly common*: for instance, one of the sacred hills north-west of Cusco is precisely called Picchu. Further, it is at least possible that the place was located in the same area of Machu Picchu but down in the valley, along the course of the Urubamba river, to which the administration of agricultural activities referred. Finally, it cannot be forgotten that there is no archaeological evidence for the presence of Spaniards in the town at any time (in spite of existing claims, also no signs of intentional destruction of Inca "idols" - a fact exceedingly common in all the Cusco huacas - are visible).

In any case, also if we admit that Machu Picchu was called Picchu and therefore mentioned in these writings, we are anyhow very far from having any interpretation of the town, and the idea of the "royal estate" has to be justified by other means. To this aim attention is called on the fact that a list of agricultural terrains attributes the farmed lands of the Urubamba valley between Ollantaytambo and Chaullay to the private possession of their conqueror, the Inca Pachacuti. Picho is *not* explicitly mentioned in this context, but "since the terrains of the valley bottom belonged to Pachacuti, it is quite probable that the places at higher quotes in the same zone were part of the royal estate of the king as well" (Rowe 1990, transl. by the author). Again, however, even if admit the validity of this second implication - and of course, we are not obliged to do this - the fact that the citadel was part of the royal properties does not show that it was used as a personal estate of the Inca. This too is, therefore, an inference, and indeed Rowe cautiously concludes "we can suppose [*podemos suponer*] that the Inca ruler choose Machu Picchu as a personal estate and as a memorial of war campaigns in the zone of Vitcos" (transl. by the author). In spite of such a correct prudence, this paper is commonly considered (perhaps without having read it) as giving conclusive proofs for the interpretation of Machu Picchu as a Pachacuti's Royal estate. Instead, it is clear that this is only a sort of vague possibility based on as much as *three inter-depending implications* [Machu Picchu was called Picho -> Picho was a property of Pachacuti -> Pachacuti built it as his personal royal estate] *none of which is securely proved*.

Of course, together with the written sources, it has to be analyzed if archaeological evidences exist to support the theory. As far as the monumental architecture of the town and its relationship with the landscape and the sky are concerned, this analysis is fully developed in the main body of the present paper. Here I will instead mention some recent studies which can be of relevance for the interpretation of the town (see Burger 2004 and references therein). One is the re-analysis of the bones found in the Bingham expedition. As already mentioned, the percentage of female bones was originally over-estimated; however, a 3:2 excess of female bones remains. It is known that the Incas established a group of women in the sanctuary of the Island of the Sun whose specific role was to serve the sanctuary; perhaps the same might have happened in Machu Picchu. The re-analysis of human remains has also shown that the inhabitants (which should have been around 750 in number) exhibited an high degree of population diversity. They were therefore coming from the different parts of the empire; likewise, a significant part of the pottery recovered in tombs comes from distant provinces. This is relevant for our discussion because it is known that royal estates were staffed by *mitima,* retainers of different ethnic groups removed from their original villages. However, although valid, this observation cannot be used as a proof for the "royal estate" theory, since the very same thing holds for many other Inca state projects and in particular, again, for the sanctuary on the Island of the Sun. Finally, another relevant recent analysis is the reconstruction of the ancient climate at Machu Picchu, which turned out to have been quite similar to that of today: warmer than in Cusco, but also more rainy. It has to be definitively excluded, therefore, that Machu Picchu could have been considered as a more pleasant place to stay with respect to the capital during the rainy season (October to April), while a relative improvement of the climate temperature might actually have been obtained by moving to Machu Picchu in winter.

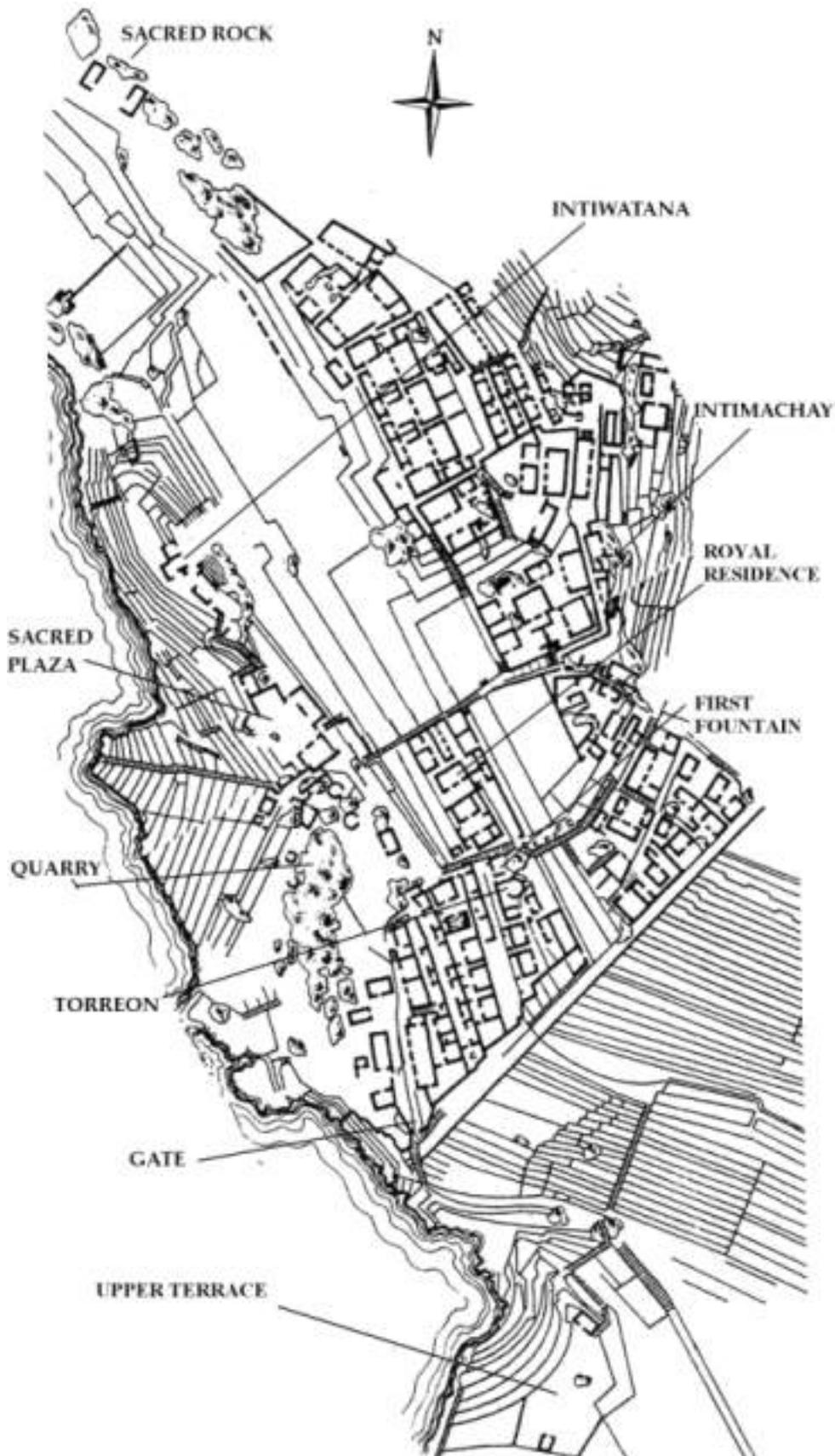

Fig. 1 North-on-top Map of Machu Picchu with highlighting of the sites discussed in the text

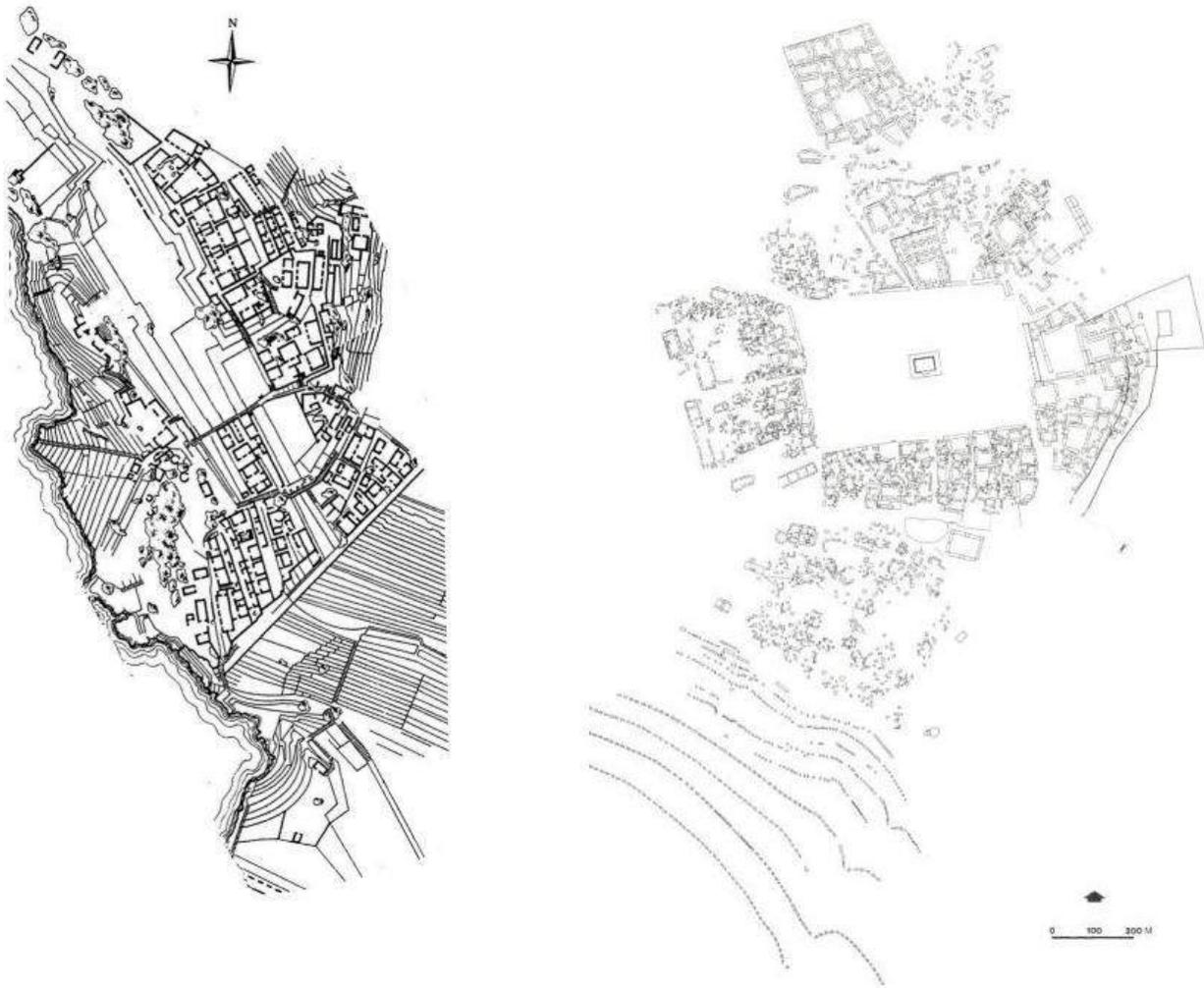

Fig. 2 Comparison between north-on-top maps of Machu Picchu and Huanuco Pampa (maps not at the same scale).

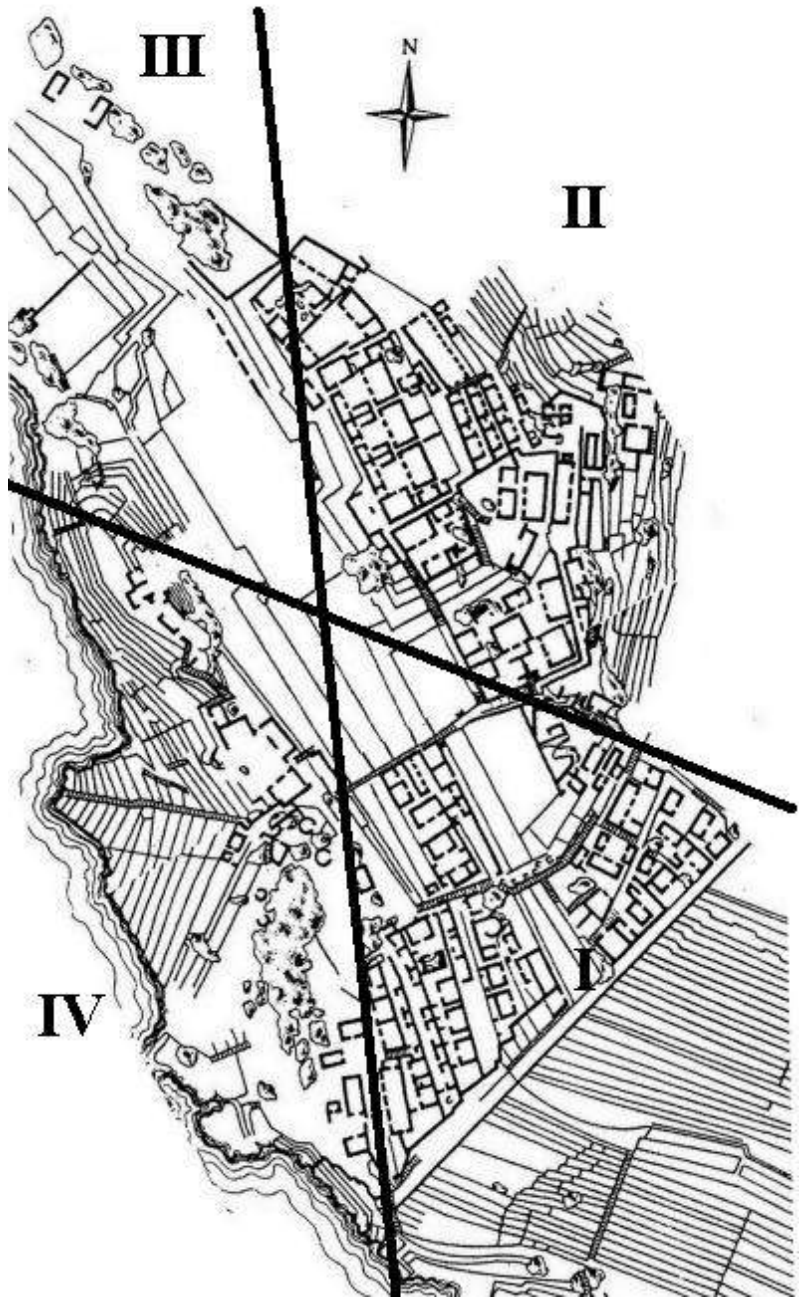

Fig. 3 The proposed "quadripartition" of Machu Picchu, see text for details.

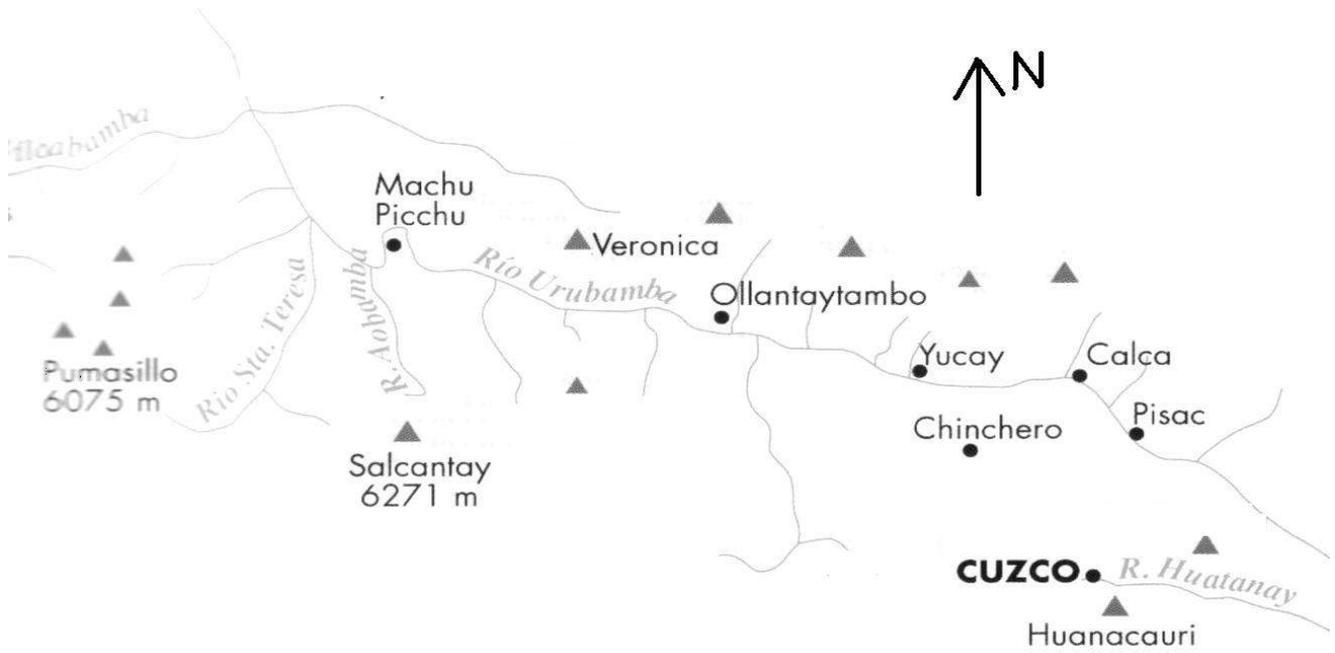

Fig. 4  The geographical position of Machu Picchu with respect to "cardinal" mountains and to the Urubamba river.

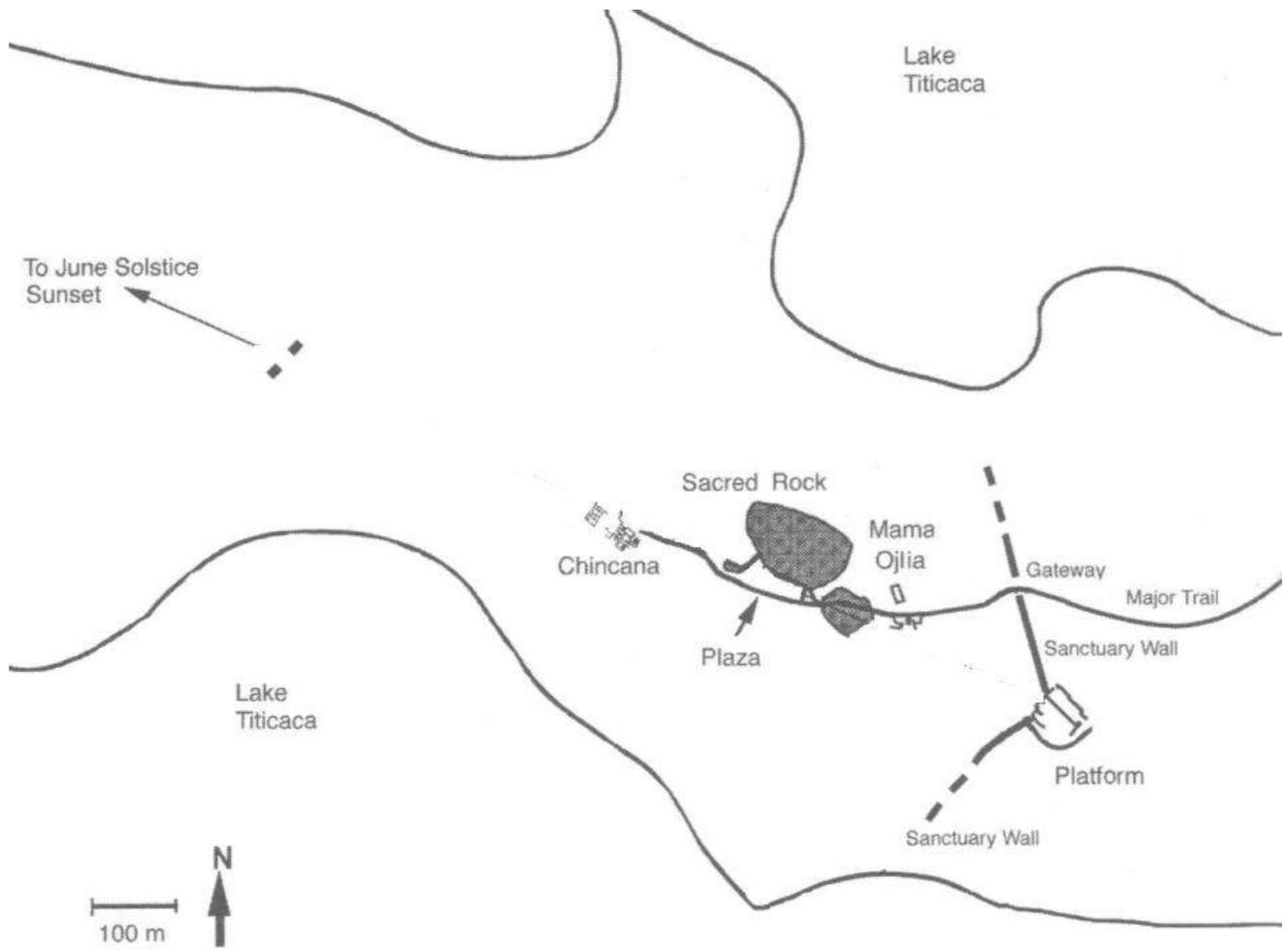

Fig. 5 Map of the Inca sanctuary on the Island of the Sun. The position of the pillars used to observe the winter solstice sunset from the sanctuary area is indicated (adapted from Bauer and Stanish 2001).

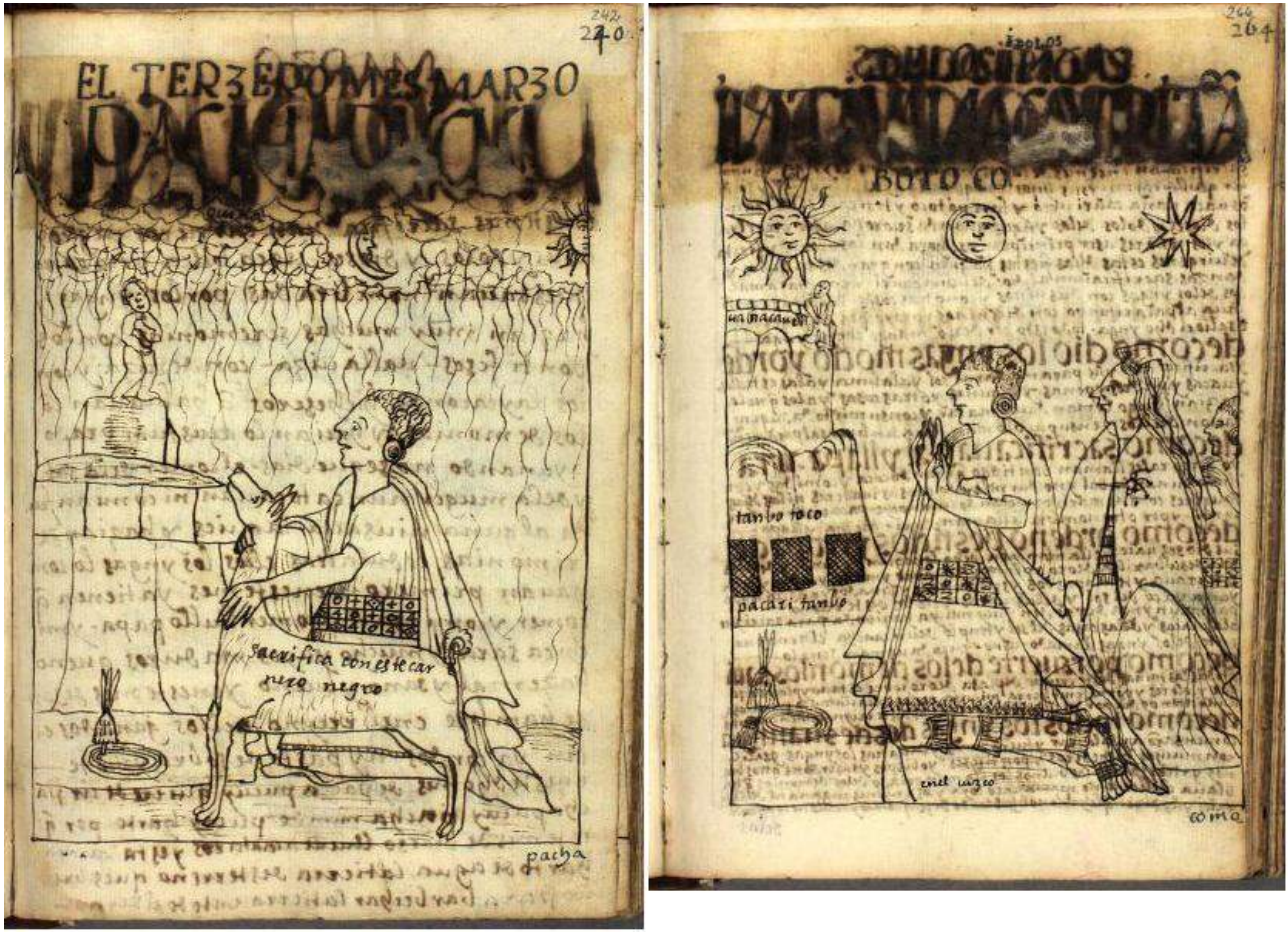

Fig. 6 Poma de Ayala *folii* 242 (right) and 266 (left).

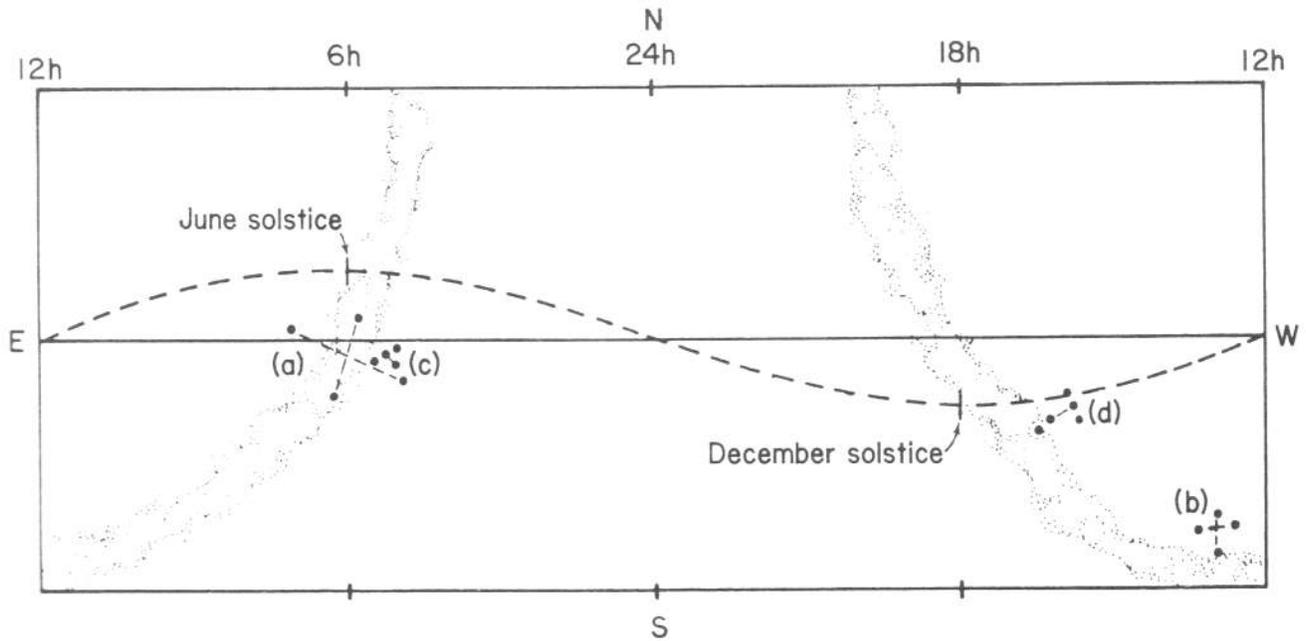

Fig. 7 The intersection of the Milky Way with the Ecliptic. The position of the solstices in the Milky Way, respectively in Sagittarius and near Gemini, is shown. Four star-to-star "cross" constellations of today's Misminay people are also shown: a) a combination of stars in the area of Gemini/Orion, b) our Southern Cross, c) The Belt of Orion plus probably Rigel, d) Five stars in the head of Scorpio (from Urton 1982).

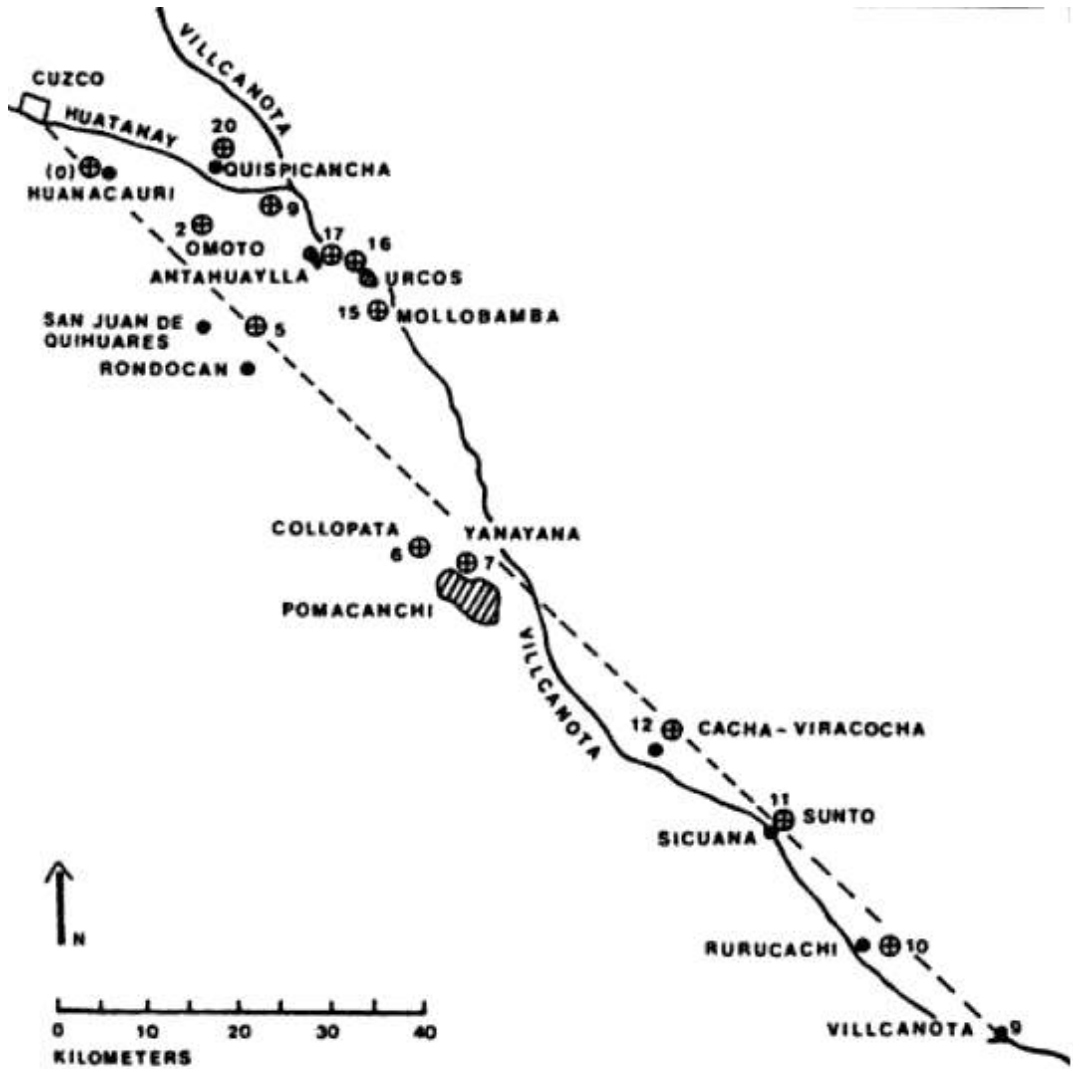

Fig. 8 The pilgrimage to the Vilcanota Hill according to Zuidema (1982).

[1] It is worth noticing that trials have been made in the past to divide the plan of the town into the two traditional moieties - the upper, or *hanan*, and the lower, or *hurin* - which are typical of the organization of the Andean space. According to such a principle was, for instance, ideally divided the capital Cusco. Contrary to Cusco, however, where this division is well documented in the chronicles and essentially corresponds to the northern and the southern parts respectively, at Machu Picchu the situation is unclear. The standard proposal is anyway to individuate the hanan part as that including the "royal residence" and the quarry (sectors I and IV in Fig. 3).

[2] The level of precision of such measurements has been the subject of much debate starting from a paper by Dearborn and Schreiber (1986). This topic is not of specific relevance here however; the interested reader can consult the anthology recently edited by A. Aveni (2008).

[3] The possible existence of other functions for the town (for instance, administrative) are not excluded *a priori*, but have to be considered as subsidiary to the main one. Similarly, although the "royal estate" theory is refused here, this of course does not mean that the Inca did not visit the town. On the contrary, it is very likely that he did and the evidences for a "royal residence" are actually quite compelling.

[4] A similar "symbolic + functional" interpretation has been proposed by who writes for the famous quarry of Ranu Raraku on Easter Island, where several huge statues were left at different stages of extraction and completion (Magli 2009).

[5] It can be noticed that important oracular shrines are documented in the Andean world since extremely ancient times, e.g. in Chavin and Pachacamac, and that some of the caves at Machu Picchu (especially the Temple of the Condor) may suggest the same interpretation.

[6] It is worth recalling that the coincidence between the solstices and the intersections of the ecliptic with the Milky Way was fundamental in Maya cosmology, see e.g. Schele, Freidel and Parker (1995)

[7] People living in the northern hemisphere associate "north" naturally with "up", because their celestial pole - and therefore the center of the apparent rotation of the stars, the place where the *axis mundi* meets the sky - is the northern one. Interestingly, it is apparent that north was associated with "upper" (hanan) also by the Incas, although they were living in the southern hemisphere. The anthropological reasons for this fully merit, at least in the author's view, further investigations.